# An efficient neural optimizer for resonant nanostructures: demonstration of highly-saturated red silicon structural color

Ronghui Lin†, Vytautas Valuckas†, Thi Thu Ha Do, Arash Nemati, Arseniy I. Kuznetsov, Jinghua Teng*, Son Tung Ha*

Institute of Materials Research and Engineering (IMRE), Agency for Science, Technology and Research (A*STAR), 2 Fusionopolis Way, Innovis #08-03, Singapore 138634, Republic of Singapore
* Corresponding author emails: jh-teng@imre.a-star.edu.sg; ha_son_tung@imre.a-star.edu.sg.
† These authors contributed equally to this work

**Abstract**
Freeform nanostructures have the potential to support complex resonances and their interactions, which are crucial for achieving desired spectral responses. However, the design optimization of such structures is nontrivial and computationally intensive. Furthermore, the current "black box" design approaches for freeform nanostructures often neglect the underlying physics. Here, we present a hybrid data-efficient neural optimizer for resonant nanostructures by combining a reinforcement learning algorithm and Powell's local optimization technique. As a case study, we design and experimentally demonstrate silicon nanostructures with a highly-saturated red color. Specifically, we achieved CIE color coordinates of (0.677, 0.304) - close to the ideal Schrödinger's red, with polarization independence, high reflectance (>85%), and a large viewing angle (i.e., up to ± 25º). The remarkable performance is attributed to underlying generalized multipolar interferences within each nanostructure rather than the collective array effects. Based on that, we were able to demonstrate pixel size down to ~400 nm, corresponding to a printing resolution of 65,000 pixels per inch. Moreover, the proposed design model requires only ~300 iterations to effectively search a 13-dimensional design space - an order of magnitude more efficient than the previously reported approaches. Our work significantly extends the freeform optical design toolbox for high-performance flat-optical components and metadevices.

**Introduction**
Recent advances in metaoptics brought many breakthroughs in a wide range of applications, such as optical computing (*1–3*), signal processing (*4*, *5*), imaging (*6–8*), and communication (*9–11*). In particular, freeform structures (*12*, *13*) realized by the inverse design methods enable versatile control of the wave propagation for multifunctional (*14–16*), broadband (*17*, *18*), and polarization-controllable photonic devices(*19*). However, designing such freeform metadevices often involves optimization in high-dimensional design space. With the increase in dimensionality, the volume expands so quickly that available data become sparse. Concepts such as proximity, distance, or nearest neighbors are no longer meaningful descriptions of spatial relationships between data points (*20*). These problems, summed up as "the curse of dimensionality" (*21*), remain an open challenge despite recent impressive results in photonics inverse design (*13*, *22*), where there are two main categories of approaches.

The first category involves gradient-based methods and heuristic algorithms such as particle swamp optimization (PSO), simulated annealing (SA), and differential evolution (DE) (*12*, *13*). These algorithms take inspiration from a physical scenario to improve the target function iteratively (*23*). They usually take a long time to converge, especially for structures with large degrees of freedom (DOF). The adjoint sensitivity analysis method can obtain the gradient of the target function for each point in the design space in just two runs by virtue of the Lorentz reciprocity (*13*, *24*). However, it is a local optimization process with no guarantee of global extrema, which is exacerbated by the fact that photonics design problems are mostly nonconvex (*12*). The difficulty in adopting nonquadratic merit functions (*13*) further restricts this algorithm.

The second category consists of data-driven algorithms that rely on the statistical characteristics of the entire training data, allowing for a more comprehensive overview of the optimization problem. However, the bottleneck is to obtain sufficient high-quality data. Although classical universal approximation theorems (*25*, *26*) state that deep neural network (DNN) is capable of approximating any function and its derivatives with arbitrary accuracy, in practical scenarios, the fine structures of the target function are often lost due to the poor quality or deficiency of the training data and the interpolation error (*27*). This results in mode collapse (*28*), where the algorithm can output only a limited number of similar designs and trivial optical responses. In addition, statistical methods are unsuitable for the exploration of underlying physics and devices with exceptional performance since these designs are often found at the anomaly and out-of-distribution data points. Given the vast number of training data needed, data-driven methods are more suitable for similar design tasks with little variations of the design targets (*27*). The recent approaches combining machine learning with optimization techniques (*23*, *29–31*) typically use machine learning to provide rough value for further optimization and thus, still suffer from the restrictions of the training data.

The adoption of reinforcement learning (RL) (*32*) for inverse design marks a complete shift of paradigm. Instead of obtaining a direct mapping between the input and the output spaces, the RL technique is a process-based technique that strives to maximize a reward function iteratively in reaction to the environment. In this regard, it resembles the conventional iterative algorithms. However, the updating rules are learned by the algorithm through continuous trial and error, resulting in a long training time (*12*, *22*). Despite a vast interest in both smart machine learning and conventional optimization algorithms (*13*, *22*, *23*, *27*, *33–36*), a global algorithm that combines speed, efficiency, accuracy, diversity of design geometries, and diversity of optical response, which can explore the extrema of target space is still lacking.

In this work, we take inspiration from RL to develop an exploration type of hybrid optimization algorithm. By introducing an efficient bifurcation mechanism that can quickly identify the good and bad regions in the optimization space, our algorithm can converge at a much faster rate compared to previous approaches. Combined with a local search technique, the global extrema can be identified rapidly. As a case study, we used this algorithm to tackle an interesting phenomenon: the absence of angle-independent, high-saturation red structural color in the natural world (*37*, *38*). This phenomenon was found to be related to a universal property of waves that higher-order resonance modes always accompany their fundamental modes, resulting in reflection peaks in the blue/green regions, hence lower red saturation in structural colors (*37*, *38*). The same principle hinders the realization of high saturation artificial red structural colour (*39–43*). A class of high-performance red color, called Schrödinger's red, requires zero reflection and perfect reflection correspondingly in different spectral ranges. A recent work (*41*) uses double bound state in the continuum (BIC) modes to reduce the high-order resonance modes in the blue/green region, resulting in high saturation red color in amorphous silicon (a-Si) metasurfaces. However, due to the nature of BIC, which is based on the array effect, the red color can only be realized in a large array size (i.e., tens of micrometres) and is highly polarization and viewing-angle dependent, which would have severe limitations to some applications such as color display or printing. In this article, we show that the reflection induced by higher-order modes can be eliminated by multipolar destructive interference in a properly designed free-form resonant structure to achieve red structural color with the highest saturation ever reported.

**Results and Discussions**

To optimize the amorphous silicon (a-Si) nanostructure with a red spectral response, we consider an n-dimensional parameter space $S$, where $x \in S$ is a design element defined by the geometrical parameters. Next, a DNN model is used to map the latent representation of these parameters. A figure of merit (FOM) function $f[t(x)]$ can be defined based on the optical response $t(x)$ of the structure. In this case, the targeted optical response is Schrödinger's red spectrum. The goal of the optimization algorithms is to find $x^*$ that satisfied $x^* = arg\ max/min\ f[t(x)]$. For a generic photonics design problem, $f[t(x)]$ is typically inexplicit and little assumption can be made about its properties. However, one could draw random samples from $f[t(x)]$ to explore its property. Here we use machine learning to effectively learn the landscape information and progressively draw better samples.

Figure 1(a) illustrates the design model proposed in this work. Machine learning is used in two stages of the model: the geometry representation and the extrema searching. The geometry is represented by a 13-dimensional input array $x = [x_1, x_2, x_3 ... x_{13}]$, including the unit cell size, the nanostructure thickness, the in-plane scaling factor, and 10 dimensional vectors that can be interpreted by the variational autoencoder (VAE) into various geometrical shapes. The VAE learns an efficient compression of the training data by passing the data through a bottleneck layer, after which the bottleneck layer can be used as a latent representation of the geometrical shapes. The detailed implementation of the VAE can be found in a previous publication (*44*). An advantage of using VAE is that the variation of the input parameters $\Delta x$ results in minute variations of the geometrical shape $\Delta d$, which in turn results in a quasi-continuous variation of the optical response $t(x)$ (*44*). This is because DNNs are combinations of affine transformations and rectified linear unit (ReLU), hence can be represented by piecewise smooth tropical geometries (*45*). This property is the foundation for our subsequent optimization process. The VAE produces a wide variety of shapes with different topologies, as illustrated in Supplementary Information 1. Furthermore, the fabrication constraints are addressed automatically by the VAE because the ultrafine details of the geometrical shapes are eliminated in the data compression process. Using numerical methods, we obtain the broadband reflection spectra $t(x) = [t_1, t_2, t_3, ..., t_{100}]$, where $t_i$ is the reflection at each discrete wavelength. The FOM function $f[t(x)]$ is the coordinates in CIE 1931 RGB color space calculated from the reflection spectrum.

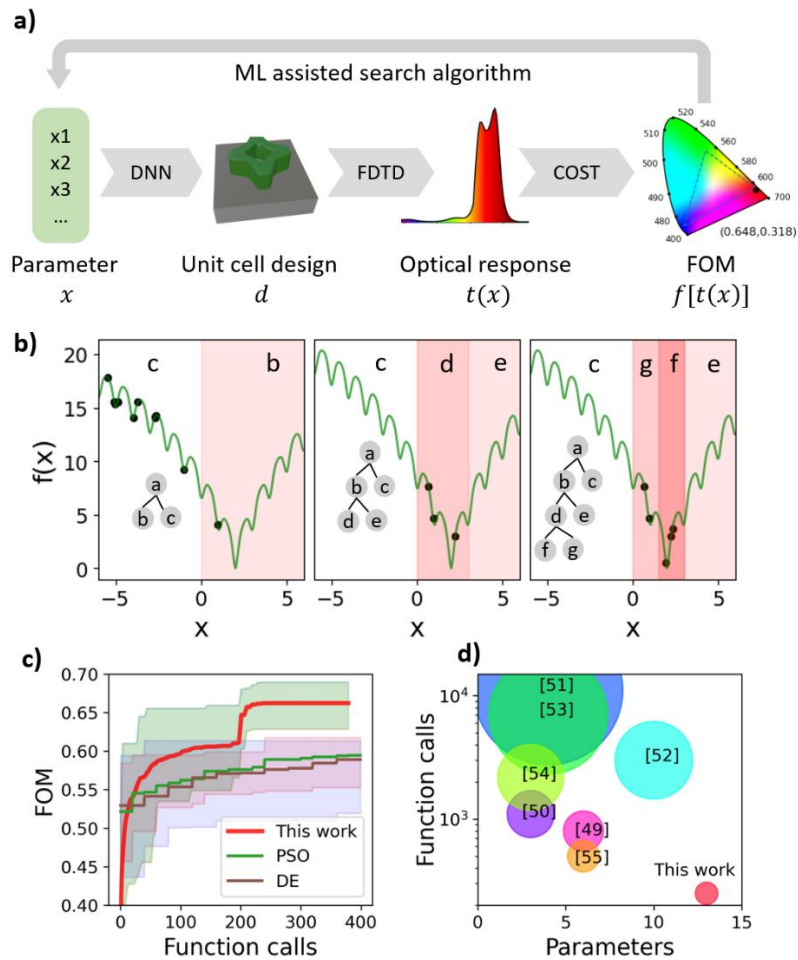

**Fig. 1**. (a) A schematic diagram of the algorithm framework. (b) Process for subdividing the search space using the MCTS algorithm. The shaded area represents the area with high FOM. (c) Comparison of the convergence speed of FOM with various conventional optimization algorithms. (d) Comparison of the algorithm efficiency with published works.

The FOM is fed to a machine learning-guided search algorithm to find the extrema. The backbone of the search algorithm is the combination of machine learning with Monte Carlo tree search (MCTS), which has been used in other large-scale artificial intelligent projects such as in AlphaGo (*46*). We adopt a variation of the MCST algorithm proposed in a previous publication (*47*). A detailed description of the MCTS algorithm can be found in Figure 1(b). We draw a few random initial samples, and the data pair $(x, f(x))$ are fed to k-means clustering algorithm to distinguish them into the good and bad regions (b and c in Figure 1(b), left panel). A decision boundary between the good and bad regions is determined by support vector machines (SVMs). After that, new samples are drawn from the good region. When the number of the samples in the good region reaches a threshold, it is further divided into two subsections (Figure 1(b) middle panel), and the process continues until the preset iterations are reached or when the FOM reaches a certain limit. Here SVMs are used for boundary determination because they converge fast and require fewer data in high dimensional spaces compared to DNNs. For the simple purpose of distinguishing good and bad regions, SVMs have proven to be data efficient. As the search tree grows, the search area is scaled down, and the estimation of the target landscape becomes more precise. After reaching a threshold of FOM or iteration numbers, Powell's method (*48*) is triggered as a local search algorithm to pin down the extrema value.

Supervised machine learning algorithms strive to obtain an approximation of the original function $f'[t(x)] \sim f[t(x)]$ through the statistics of the training data. This requires an exponential increase of training data as the dimension of $x$ increases to ensure the approximation precision. The use of machine learning as a search algorithm to explore the landscape of $f[t(x)]$ iteratively reduces the amount of the training data significantly. The proposed algorithm is benchmarked against PSO and DE algorithms, which are two widely used global optimization algorithms. Each algorithm is repeated 20 times, and their convergence is shown in Figure 1 (c). The shaded area represents the maximum and minimum FOM obtained at each function call, while the thick colored lines are the averaged FOM value. The proposed algorithm converges faster and eventually reaches a higher FOM. The kink in the red curve represents the switch from MCTS to local Powell's method. It is worth noting that the population number of PSO and DE is 15. Thus, each generation of these algorithms takes 15 function calls. Figure 1(d) compares our algorithm with previous reports (49–55) using RL and heuristics optimization algorithms. The typical number of function calls are in the $10^3$ range with a much smaller design parameter size for previous reports, as opposed to around 300 function call in 13-dimensional space in this work. The high efficiency in data usage and convergence speed is attributed to the bifurcation rule and the efficient good-region identification by SVMs in the proposed algorithm.

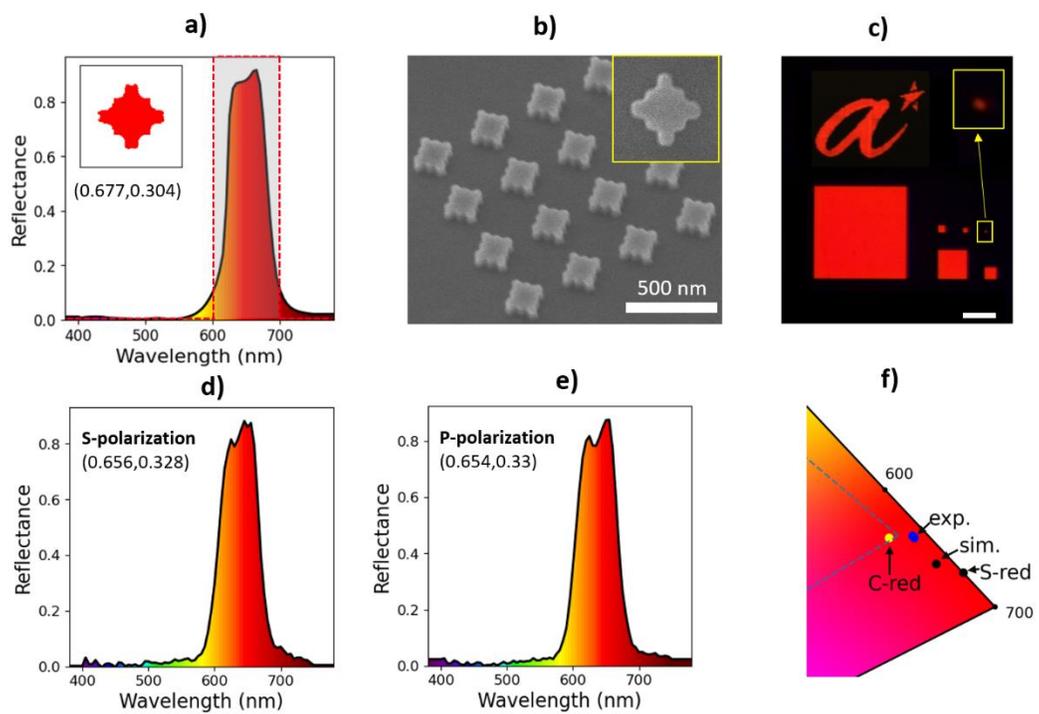

**Fig. 2.** (a) Optimized a-Si nanostructure (inset) and its simulated reflection spectrum. The shaded area shows the spectrum of the ideal Schrödinger's red color. (b) Tilted-view scanning electron microscope (SEM) image of a fabricated a-Si pixel. The inset shows a top-view SEM image of an a-Si unit cell. (c) Optical microscope image of the fabricated red pixels with various sizes (i.e., 30, 10, 5, 2.5, 1.5, and 0.4 $\mu m \times \mu m$) and A*STAR logo made of the a-Si nanostructures. The scale bar is 10 $\mu m$. The inset is a magnified optical image of the 0.4 $\mu m$ pixel. (d, e) Reflection spectra at normal incidence for s-polarization and p-polarization, respectively. (f) The CIE coordinates of the optimized color in simulation (sim.) and experiments (exp.). The ideal Schrödinger's red pixel (S-red), and Cadmium red (C-red) are marked for reference. The dashed line is the boundary of the sRGB standard.

Using the optimization process described above, we derive an a-Si nanostructure with a reflection spectrum that matches well with the ideal Schrödinger's red spectrum (i.e., our target optical response), as shown in Figure 2(a). The optimal structure has a unit cell of 393 nm and a height of 120 nm. We then fabricate the a-Si nanostructures using conventional e-beam lithography and reactive-ion etching processes. Figure 2 (b) shows the scanning electron microscope (SEM) images of the fabricated a-Si nanostructures. Detailed fabrication processes can be found in the Methods section. Figure 2(c) presents the optical microscope image of the fabricated structures showing high purity red color even for the smallest pixel made of 1 unit cell with a lateral dimension of ~400 nm, corresponding to a printing resolution of ~65,000 pixels per inch. The simulated reflection spectra for different pixel sizes can be found in Supplementary Information 2, Figure S1. The data show that the high-purity color depends weakly on the pixel size and more on the scattering of the individual elements. Figure 2(d, e) shows the measured reflection spectra for the 30 $\mu m$ pixel under s and p polarizations, respectively. It can be seen that, for both polarizations, the high reflection (i.e., > 85%) in the red region (i.e., 600 – 700nm) and the suppression of the reflection in the shorter wavelength region (i.e., 400 – 600 nm) are achieved simultaneously, which agree well with the simulated reflection spectrum shown in Figure 2(a). The corresponding coordinates in the CIE 1931 color space are (0.656, 0.328) and (0.654, 0.33) for s and p polarization, respectively, which are slightly lower than the simulated structure (0.677, 0.304). This can be explained by the discrepancy in the designed and fabricated structures (Insets to Figure 2(a, b)). Nevertheless, these CIE coordinates are among the best red pixel ever reported (*41*). Furthermore, the color is polarization independent owing to the symmetry of the design. Figure 2(f) shows the color space coordinates of the a-Si nanostructures in comparison with the ideal Schrödinger's red (S-red) and Cadmium red (C-red). It can be seen that the experimental value for a-Si nanostructure in this work surpasses the pigment Cadmium red in terms of color purity, and the simulated value approaches that of the S-red. With further improvement in nanofabrication to replicate the smaller features in the design, higher saturation of the red color can be achieved.

The angle-resolved reflection spectra are measured to investigate the performance of the a-Si red pixel at a wider field of view, as shown in Figure 3 (a, b) for simulation and experiment, respectively. For p-polarization, the red reflection band remains up to ± 18° viewing angle, while for s-polarization, the high reflection band persists up to ± 26°. Remarkably, the suppression of reflection at shorter wavelengths (i.e., 400-600 nm) can be maintained over the whole measurement angle range (i.e., ± 26º). This can be attributed to the generalized multipolar interaction, which ensures the cancellation of outgoing waves at wider angles. The case for unpolarized light can be viewed as the superposition of p-and s-polarized light. The narrower viewing angle for p-polarization can be explained by the lower efficiency to excite the horizontal electric dipole at higher incident angles. The normalized E-field distributions within the structure at 650 nm (i.e., red region) and 550 nm (i.e., green region) are shown in Figure 3 (c, d), respectively. It can be clearly seen that the field is much stronger for the case of 650 nm compared to that of 550 nm, indicating the suppression of resonance at lower wavelength region.

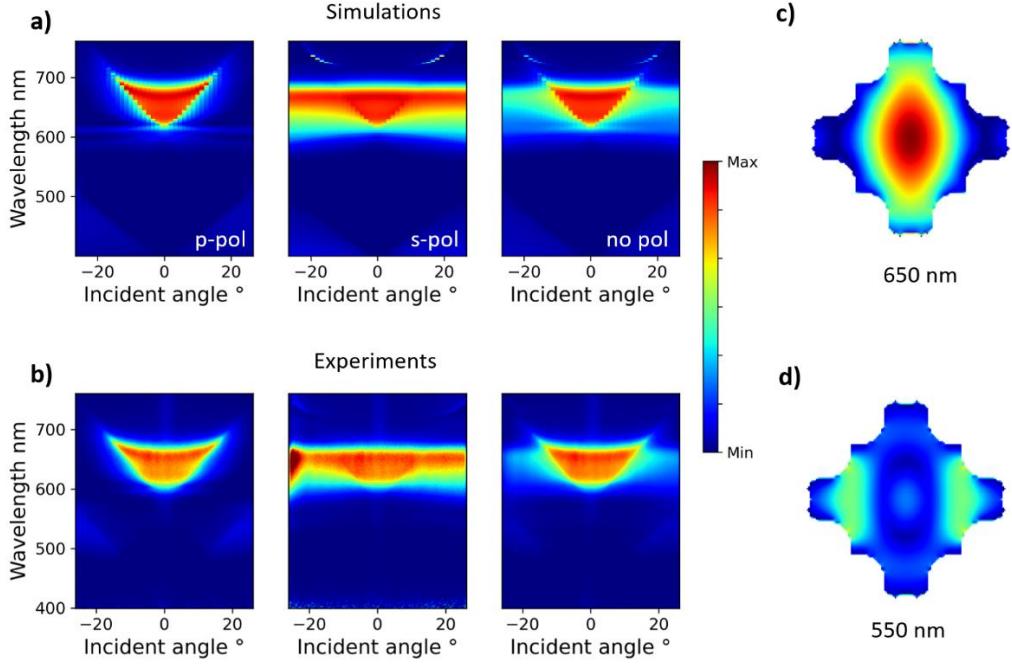

**Fig. 3.** Simulated (a) and measured (b) angle-resolved reflection spectra for 30-$\mu m$ pixel. (c, d) Normalized E-field intensity distribution inside the a-Si nanostructure at 650 nm and 550 nm, respectively.

To shed more light on the optical responses of the a-Si color structure, we studied the multipolar contributions to the scattering based on the induced electric current density. This can be expressed as $j(r) = -i\omega\varepsilon_0(n^2 - 1)E(r)$, where $\omega$ is the angular frequency, $\varepsilon_0$ is the vacuum permittivity, and $E(r)$ is the electric field distribution within the unit cell. The induced current density is expanded into electric dipole (ED) $p$, magnetic dipole (MD) $m$, electric quadrupole (EQ) $Q$, magnetic quadrupole (MQ) $M$ and electrical octupole (EO) $O$ in cartesian coordinates using the exact multipole expansion formulation (*56–58*). The reflection coefficient for x polarized incident plane wave can be obtained from the multipole moments using the following equation (*57*):

$$r = \frac{ik_d}{2E_0 S_L \varepsilon_0 \varepsilon_d}\left(p_x - \frac{1}{v_d}m_y + \frac{ik_d}{6}Q_{xz} - \frac{ik_d}{2v_d}M_{yz} - \frac{k_d^2}{6}O_{xzz}\right)$$

Here, $k_d = k_0\sqrt{\varepsilon_d}$ is the wave number in the surrounding medium, $E_0$ is the incident electric fields, $S_L = D^2$, where $D$ is the lattice constant for the square lattice. $\varepsilon_0$ and $\varepsilon_d$ are the vacuum and relative permittivity of the surrounding medium, respectively. The total reflection is calculated as $R = |r|^2$. To visualize the multipole interaction, we factor in all coefficients and represent the reflection coefficient as follows:

$$r = |C|\left(A_p e^{i\varphi_p} + A_m e^{i\varphi_m} + A_{Q_e} e^{i\varphi_{Qe}} + A_{Q_M} e^{i\varphi_{Qm}} + A_{O_e} e^{i\varphi_{Oe}}\right)$$

Where $A_i$ and $\varphi_i$ (with $i = p, m, Q_e, Q_m, O_e$) are the magnitude and phase of each multipole component, respectively.

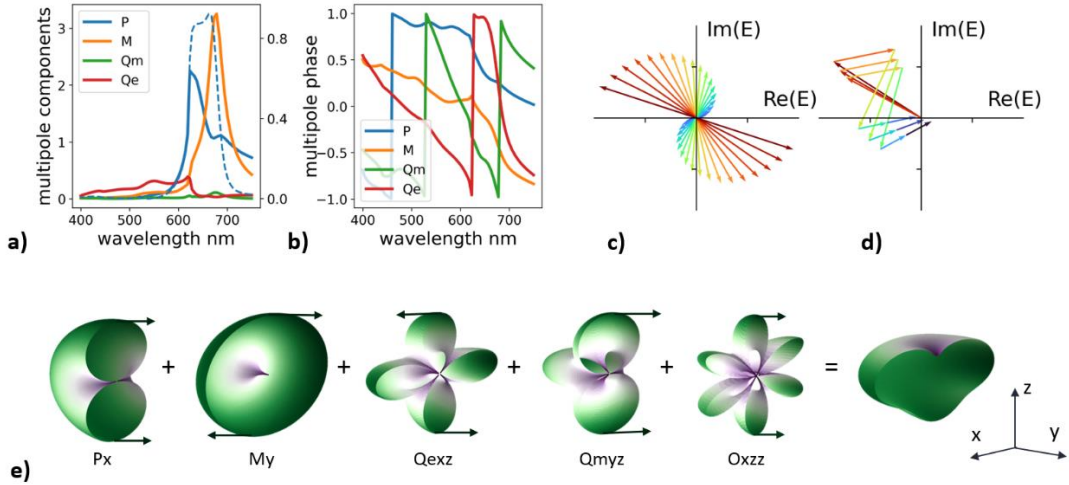

**Fig. 4**. (a) Multipolar decomposition for the optimized reflection spectra in a-Si nanostructure; The solid lines are multipole magnitude, and the dashed line is the reflection spectra. (b) The phase spectra of the multipoles in (a). (c) The cancellation condition for the case of two-multipole interference. (d) The cancellation condition for multiple components interference at three different wavelengths: 548 nm, 549 nm, and 550 nm. Each arrow represents one multipolar component. (e) Far-field distribution of the multipolar components and their interference pattern.

As shown in Figure 4 (a, b), the electric dipole and magnetic dipole responses dominate the high reflection band in the red region (i.e., >600 nm), while the EQ mode is dominant at the shorter wavelength region (i.e., <600 nm). In the optimized structure, the magnitude of EQ response is low and is further suppressed by the interaction with other multipolar components. Figure 4 (c) shows the destructive interference conditions for the case of two multipoles which generally require opposite phases and the same magnitude. Since each multipolar component has different wavelength dependence, it is challenging to achieve broadband cancellation for two multipoles. However, it is possible in the case of multiple multipole interactions. Figure 4(d) shows that the cancellation condition for three different wavelengths can be achieved by varying the phase and amplitude of each multipole component. Furthermore, the cancellation also guarantees high suppression of the reflected field at a wide angle, as shown in the far field distribution in Figure 4(e). We further discuss the tuning power of the multipoles by comparing the optimized structure with various sub-optimized ones in Supplementary Information 2, Section 3 & Figures S2-S4. The result shows that the small variation of the geometries can effectively tune the multipole moments and phases to reach the cancellation condition at a shorter wavelength and enhancement condition at a longer wavelength. Controlling of two multipoles components is critical in many fascinating phenomena such as Fano resonances (*59*), electromagnetically-induced-transparency (*60*), unidirectional scattering (*61*, *62*), Janus and Huygens dipoles (*63*), and perfect absorption (*64*), to name just a few. They originate from the different parity and polarization signatures of each multipole radiation pattern. However, previous reports rely on two multipole components with a specific phase and amplitude relationship at a specific wavelength. The freeform structures allow for more general interaction involving multiple multipoles with generic phase and intensity relationships, providing a possibility for other wide bandwidth photonic devices.

**Discussion**

We designed and experimentally demonstrated a-Si structural red color with a record-high performance by employing an efficient data-driven machine learning approach. The optimal free-form structure enables the elimination of high-order resonances in the a-Si structure due to their destructive interferences, leading to the highest ever reported saturation of red color with a CIE coordinate of (0.677, 0.304) in simulation and ~(0.656, 0.328) in experiments. Remarkably, the color structures have low polarization dependence and high viewing angle (i.e., up to 26º), which are unprecedented for such high saturation color (*41*). The introduction of a simple bifurcation rule and the use of SVMs in determining good and bad optimization regions are the keys to rapid convergence. The design model proposed in this work significantly advances the free-form design approach for high-performance resonant nanostructures, which may find broad applications in color filters, spectrometers, and sensors.

## Materials and Methods

### Numerical simulation

We use commercial software Lumerical® FDTD to obtain the reflection spectrum of the Si structures by assuming periodic boundary conditions. The machine learning algorithm is implemented in Python using the Pytorch open package. The refractive index of a-Si is based on our ellipsometry measurements (*65*). The details for multipolar expansion can be found in Supplementary Information 2, Section 1.

### Nanofabrication of Si color pixels

Amorphous silicon film was grown onto a thin quartz substrate using chemical vapor deposition (CVD, OPIT Plasmalab 380). The thickness of the deposited film is then characterized using a reflectometer (Filmetrics® F20). The measured thickness of a-Si film is ~123 nm, which is close to the designed value of 120 nm. The sample was then patterned using a standard electron-beam lithography (EBL) process: First, hydrogen silsesquioxane (HSQ) resist (Dow Corning, XR1541-002) is spin-coated on the a-Si/quartz substrate at 3000 RPM, giving a ~150 nm-thick film. A charge dissipation layer (Espacer 300AX01) is used to minimize the charging effect. The nanopatterning is done using the Elionix ELS-7000 system at 100 kV and 500 pA. The EBL pattern is then developed using a home-made salty developer (*66*). Finally, the nanopatterns are transferred to the a-Si layer by a reactive-ion-etching process (OPIT Plasmalab 100) using HBr and $O_2$ gases.

### Optical characterizations

Angle-resolved reflection measurements of the Si color pixels were conducted using a home-built back-focal plane setup consisting of an inverted microscopy (Nikon Ti-U) and a spectrograph (Andor SR-303i) coupled with an Electron Multiplying Charge-Coupled Detector (EMCCD) (Andor Newton 971, 400×1600 pixels). A detailed schematic of the setup is described in our previous work (*67*). A 50× objective (Nikon) with a numerical aperture (NA) of 0.45, corresponding to a viewing angle of ± 26.7º, was used for illumination and collection of the reflection spectra. A pinhole was inserted into the optical path at an image plane to limit the collection area before projecting to the spectrometer slit. Reflected light was normalized to light reflected from a silver mirror in the same measuring configuration after accounting for photodetector noise effects (dark current subtraction).

**Acknowledgments**
S.T.H, V.V, T.T.H.D, and A.I.K acknowledge support from A*STAR MTC-Programmatic Fund (M21J9b0085). S.T.H and T.T.H.D also acknowledge the financial support from AME Yong Individual Research Grant (A2084c0177). J.T. acknowledges the National Research Foundation, Singapore under NRF-CRP (NRF-CRP26-2021-0004) and AME IRG Grant (A20E5c0084 and A2083c0058).

**Funding:**

AME Yong Individual Research Grant (A2084c0177)
A*STAR MTC-Programmatic Fund (M21J9b0085)
AME IRG Grant (A20E5c0084 and A2083c0058)
National Research Foundation NRF-CRP (NRF-CRP26-2021-0004)

**Author contributions**

S.T.H and R.L conceived the idea of the project. R.L developed the machine learning design model with the inputs from S.T.H. V.V fabricated the samples. T.T.H.D and S.T.H characterized the samples. A.N, A.I.K, and J.T contributed to the discussion and development of the concepts. A.N helped with the FDTD simulations. S.T.H and J.T supervise the project. R.L and S.T.H wrote the manuscript with inputs from all authors. †R.L and V.V contributed equally.

**Competing interests:** Authors declare that they have no competing interests.

**Data and materials availability:** All raw data are available upon reasonable request from the corresponding author.

**Supplementary Materials**

Supplementary Information 1: Animation showing the gradual transition of the geometry as a function of input parameters (file type, tif).

Supplementary Information 2: Multipole expansion; size-dependent reflection spectra; the multipolar decomposition of unoptimized structures; the comparison of field distribution for optimized and unoptimized structures; the SEM image and reflection spectra of sub-optimized structure. (file type PDF)

**Supplementary Information 2:**

1. **Multipole expansion**

   The multipole expansion is carried out in Cartesian coordinates based on the displacement currents $\mathbf{j} = i\omega\varepsilon_0(\varepsilon_r - 1)\mathbf{E}$, where $\omega$ is the angular frequency, $\varepsilon_0$ is the permittivity in vacuum, and $\varepsilon_r$ is the complex relative permittivity. The multipole moments are calculated as follows:

   Electric dipole moment: $\mathbf{P} = \frac{1}{i\omega}\int \mathbf{j}\, d^3r$

   Magnetic dipole moment: $\mathbf{M} = \frac{1}{2c}\int (\mathbf{r} \times \mathbf{j})\, d^3r$

   Toroidal dipole moment: $\mathbf{T} = \frac{1}{10c}\int [(\mathbf{r} \cdot \mathbf{j})\mathbf{r} - 2r^2 \mathbf{j}]\, d^3r$

   Electric quadrupole moment: $Q_{\alpha\beta} = \frac{1}{i\omega}\int [r_\alpha j_\beta + r_\beta j_\alpha - \frac{2}{3}\delta_{\alpha\beta}(\mathbf{r}\cdot\mathbf{j})]\, d^3r$

   Magnetic quadrupole moment: $M_{\alpha\beta} = \frac{1}{3c}\int [(\mathbf{r}\times\mathbf{j})_\alpha r_\beta + (\mathbf{r}\times\mathbf{j})_\beta r_\alpha]\, d^3r$

   The total radiation power of each multipole moment is:

   $$I = \frac{2\omega^4}{3c^3}|\mathbf{P}|^2 + \frac{2\omega^4}{3c^3}|\mathbf{M}|^2 + \frac{4\omega^5}{3c^4}|\mathbf{P}\cdot\mathbf{T}|^2 + \frac{2\omega^6}{3c^5}|\mathbf{T}|^2 + \frac{\omega^6}{5c^5}\sum |Q^e_{\alpha\beta}|^2 + \frac{\omega^6}{40c^5}\sum |Q^m_{\alpha\beta}|^2 + o(\frac{1}{c^5})$$

   Here, $c$ is the speed of light. α, β, γ = x, y, z are the Cartesian coordinates.

2. **Size dependent reflection**

   Figure S1(a) depicts the reflection spectra for different array size. The suppression of reflection at wavelength below 600 nm is consistent for all the array size. The reflection peak in the red band due to the response is also present for the smallest array simulated. The dropped reflection when the array gets smaller is due to the limitation of the monitor size during the simulation, which can only capture part of the reflected light. As the array size get smaller, the scattered field resembles more spherical wave, and only a small portion of the scattered field can be captured by the field monitor. However, it is clear from the measured optical microscope data, even a single unit cell shows high saturation red color. Figure S1(b) shows the corresponding CIE colors for various array size. As the array size gets bigger, the coordinates move towards the red region. It is estimated that a color saturation comparable to Cadmium red can be obtained with an array size of 6 by 6-unit cells.

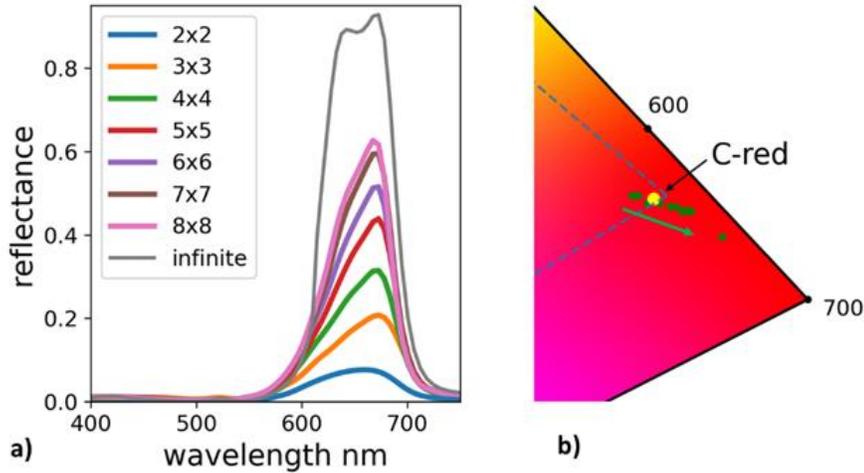

**Fig. S1.** (a) The simulated reflection spectra for different array size. (b) The green dots show the CIE coordinates for the array of different size. The Cadmium red (C-red), is also shown for reference. The dashed line is the boundary of sRGB standard. The green arrow indicates the increasing array size.

### 3. Comparison of the optimized structure with sub-optimized structures

In this section, we show the comparison of the optimized structure with various unoptimized structure to illustrate the role of the multipolar cancellation. In all the structures shown in Figure S2, the volume is conserved while only the arrangement of the material is altered. In all these structures, the reflection peak at the red region remains high. However, there is substantial reflection in the shorter wavelength region for the unoptimized structures, as shown in Figure S2(a, b). The reflection at shorter wavelength results in the CIE coordinate moving toward the blue and yellow region and therefore lower red saturation. The effective cancellation of the reflection at lower wavelength is due to the minute variation of the multipole moments and multipole phase as shown in Figure S2, last two columns to the right.

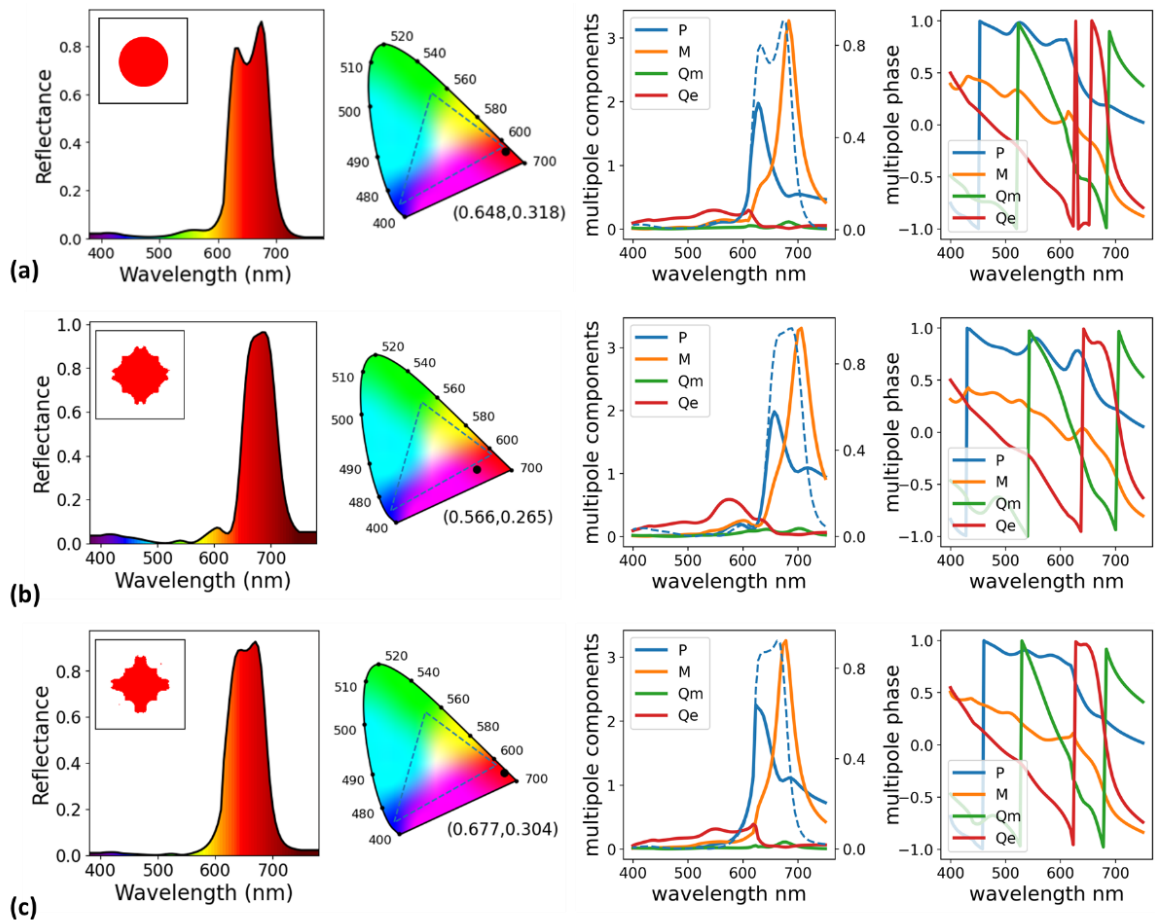

**Fig. S2.** Comparison of different geometries. (a) Circular geometry with the same volume as the optimized geometry. (b) Complex geometry with slight variation. (c) The optimized geometry. The rows in each panel are simulated reflection spectra, CIE color coordinate, multipole moments and multipole phase respectively.

The electric field intensity distribution is very similar at longer wavelength (680 nm) for the optimized structures and a slightly altered one. As the wavelength get shorter, the difference become more significant, which is also consistent with the previous observation. This further confirms the variation of the geometries can effectively modulate the optical response in a broadband wavelength.

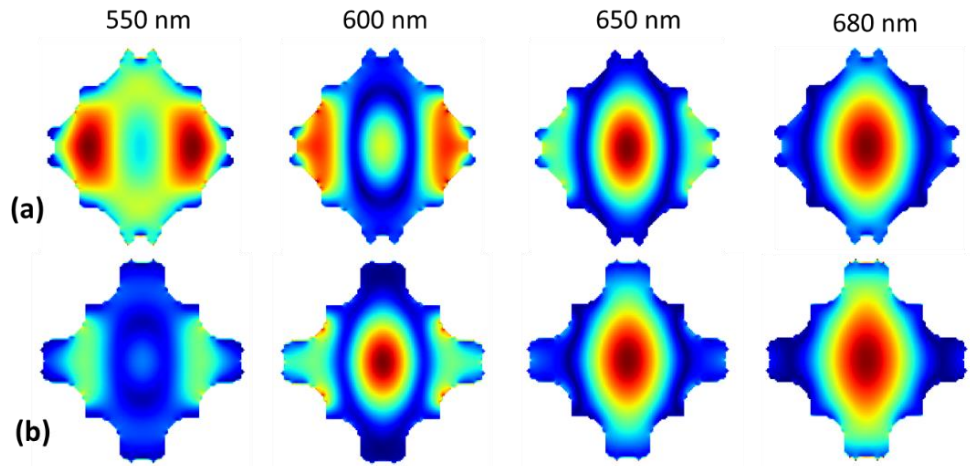

**Fig. S3.** Comparison of the E field distribution at different wavelength between (a) sub-optimized structure and (b) optimized structure.

Figure S4 (a, d) shows the comparison of the SEM images of the optimized structure and the sub-optimized one, respectively. The differences in the geometry are subtle, but the optical response is quite different, especially at shorter wavelength. Figure S4(e) shows the measured reflection spectrum of the sub-optimized structure. It can be seen the reflection in the shorter wavelength is much higher than the optimized structure, resulting in a CIE coordinate of (0.645,0.326) which is worse than the optimized one as shown in Figure S4 (c, f).

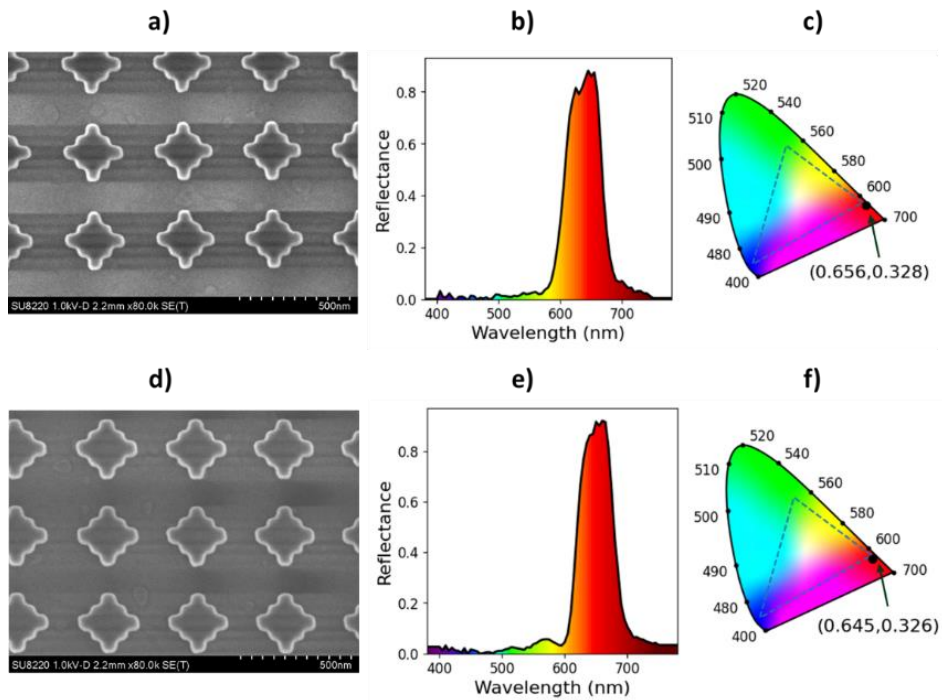

**Fig. S4.** SEM image, reflection spectrum, and CIE coordinate of the optimized structure (a-c) and the sub-optimized structure (d-f), respectively